\documentclass{webofc}

\usepackage[varg]{txfonts}   
\usepackage{hyperref}
\usepackage{url}
\hypersetup{colorlinks=true,citecolor=blue,urlcolor=blue,linkcolor=blue}
\newcommand{\jpsim}{\mathrm{J}/\psi}

\usepackage[dvipsnames]{xcolor}

\begin{document}
\title{Nuclear suppression in diffractive vector meson production within the color glass condensate framework}
%
%

\author{\firstname{Heikki} \lastname{M\"antysaari}\inst{1,2} \and
        \firstname{Hendrik} \lastname{Roch}\inst{3}\fnsep\thanks{\email{Hendrik.Roch@wayne.edu}} \and
        \firstname{Farid} \lastname{Salazar}\inst{4,5,6,7} \and
        \firstname{Bj\"orn} \lastname{Schenke}\inst{6} \and
        \firstname{Chun} \lastname{Shen}\inst{3} \and
        \firstname{Wenbin} \lastname{Zhao}\inst{8,9}
}

\institute{
    Department of Physics, University of Jyv\"askyl\"a, P.O. Box 35, 40014 University of Jyv\"askyl\"a, Finland
\and
    Helsinki Institute of Physics, P.O. Box 64, 00014 University of Helsinki, Finland 
\and
    Department of Physics and Astronomy, Wayne State University, Detroit, Michigan 48201, USA
\and
    Department of Physics, Temple University, Philadelphia, Pennsylvania 19122, USA
\and
    RIKEN-BNL Research Center, Brookhaven National Laboratory, Upton, New York 11973, USA
\and
    Physics Department, Brookhaven National Laboratory, Upton, New York 11973, USA
\and
    Institute for Nuclear Theory, University of Washington, Seattle, Washington 98195, USA
\and
    Nuclear Science Division, Lawrence Berkeley National Laboratory, Berkeley,
California 94720, USA
\and
    Physics Department, University of California, Berkeley, California 94720, USA
}

\abstract{
We perform a global Bayesian analysis of diffractive $\mathrm{J}/\psi$ production in $\gamma+p$ and $\gamma+\mathrm{Pb}$ collisions within a Color Glass Condensate based framework. 
Using data from HERA and the LHC, we find that a simultaneous description of $\gamma+p$ and $\gamma+\mathrm{Pb}$ observables is challenging. 
Introducing a global $K$-factor to account for theoretical uncertainties improves the agreement with data and enhances the framework's predictive power. 
We present predictions for integrated $\mathrm{J}/\psi$ cross sections at different photon–nucleus energies and study their $A$-dependence relative to a no-saturation baseline, quantifying nuclear suppression and providing insights into the onset of saturation effects.
}
\maketitle
\section{Introduction}
\label{intro}
At small parton momentum fractions $x$, gluon densities predicted by linear QCD evolution grow rapidly until non-linear recombination effects set in~\cite{Gribov:1983ivg}, leading to the high-occupancy Color Glass Condensate (CGC) regime~\cite{McLerran:1993ni,McLerran:1993ka}.
Diffractive vector meson production is a clean probe of small-$x$ gluons: its cross section scales (at leading order) with the square of the gluon density~\cite{Ryskin:1992ui} and is sensitive to the target's spatial structure~\cite{Mantysaari:2020axf}.
Comparing proton and nuclear targets reveals nuclear shadowing~\cite{Eskola:2022vpi} or gluon saturation~\cite{Lappi:2013am,Mantysaari:2022sux}.

CGC-based calculations constrained by $\gamma+p$ data describe both coherent and incoherent diffractive $\jpsim$ production at HERA~\cite{Mantysaari:2016ykx}, where the incoherent channel revealed the importance of a fluctuating proton substructure~\cite{Mantysaari:2018zdd}. 
Extending the same framework to nuclei in ultraperipheral heavy-ion collisions, experiments have measured $W$-dependent coherent $\jpsim$ cross sections for $\gamma+\mathrm{Pb}$~\cite{ALICE:2023jgu,CMS:2023snh}. 
These data show stronger nuclear suppression than expected, as CGC predictions constrained by HERA data typically rise more rapidly with $W$.

Here, we analyze the $A$-dependence of coherent $\jpsim$ production with a Bayesian framework that confronts both $\gamma+p$ and $\gamma+\mathrm{Pb}$ data. 
To absorb model uncertainties --- such as those from the vector meson wave function or higher-order effects --- we introduce an overall $K$-factor, determined alongside the other parameters. 
Comparing to the no-saturation limit, we quantify the nuclear suppression and assess its connection to saturation dynamics.

\section{Model}
To compute the diffractive $\mathrm{J}/\psi$ production cross section, we employ the CGC framework including the JIMWLK energy evolution used in Ref.~\cite{Mantysaari:2025ltq}.
In this work, we focus on coherent cross-section which, as a function of Mandelstam $t$, reads
\begin{equation}
\label{eq:coh}
    \frac{\mathrm{d}\sigma^{\gamma + A \to \jpsim + A}}{\mathrm{d}t} = \frac{K}{16\pi} \left| \left\langle \mathcal{A}^{\gamma^*+p \to V + p} \right\rangle_\Omega  \right|^2.
\end{equation}
Here $\langle \mathcal{O}\rangle_\Omega$ refers to an average over target configurations $\Omega$, and $\mathcal{A}$ is the amplitude computed from the wave function overlap of the photon and the vector meson, together with the dipole-target amplitude obtained from the Wilson lines in the McLerran-Venguopalan model.
The energy evolution of the Wilson lines going into this expression is computed using the JIMWLK evolution on an event-by-event basis~\cite{Mueller:2001uk}.
In Ref.~\cite{Mantysaari:2025ltq}, it was explored whether additional parameters, such as an additional proton shape parameter or an additional high-frequency regulator for the Wilson lines, should extend the model.
It turns out that the combined data from a Bayesian fit to the $\gamma+p$ and $\gamma+\mathrm{Pb}$ data disfavors most extensions of the model with additional parameters, but we find that the addition of a $K$-factor -- scaling all cross sections by a constant -- is favored.
We make use of this minimally extended model and the generated posterior distribution from Ref.~\cite{Mantysaari:2025ltq} to make predictions using 25 parameter samples from the posterior distribution, allowing us to propagate the variation of the parameters in the posterior distribution to the final predictions.
In Ref.~\cite{Mantysaari:2025ltq}, the preferred value of the overall $K$-factor was found to be $\sim 0.33$, implying a strong rescaling of the cross-section normalization. 
A value $K<1$ is compensated in the fits by a larger color charge density, which corresponds to denser nucleons and, in turn, stronger nuclear suppression.

\section{Results}
We make predictions using the same computational setup as in Ref.~\cite{Mantysaari:2025ltq} for nuclei ranging from the proton up to Uranium at two center-of-mass energies, $W=31.5$ and $813$~GeV, corresponding to momentum fractions $x=0.01$ and $1.5\times 10^{-5}$ respectively. This allows us to study how the nuclear suppression of the coherent cross section develops relative to the $\gamma+p$ case.
\begin{figure}[!htb]
\centering
\sidecaption
\includegraphics[width=8cm,clip]{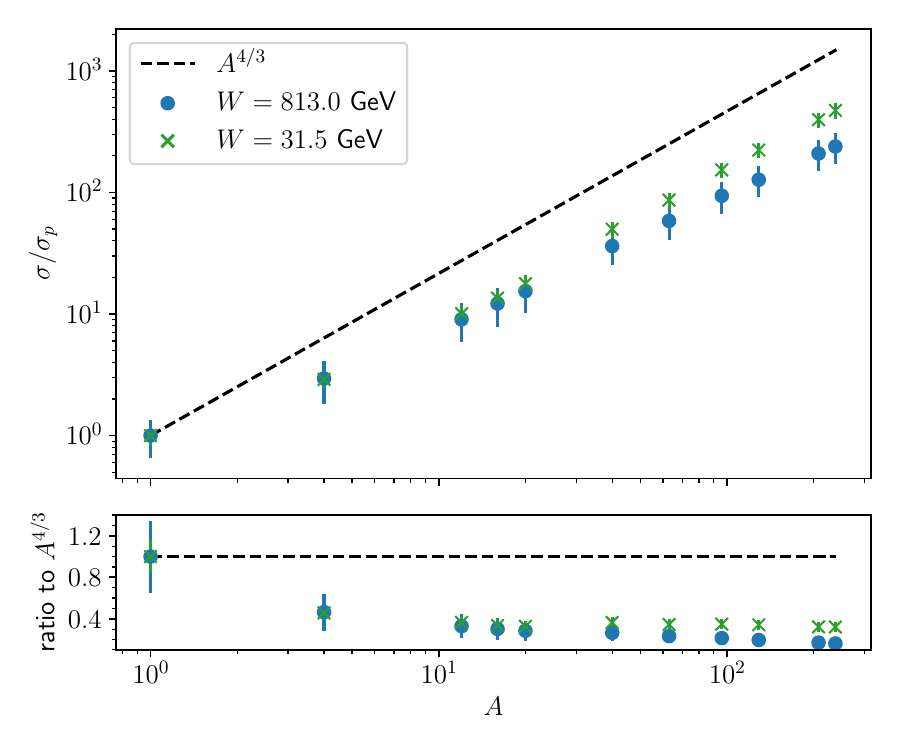}
\caption{Ratio of the integrated coherent $J/\Psi$ cross section to the $\gamma+p$ case as a function of the nuclear mass number $A$ for $W=31.5$ and $813$~GeV. The dashed line shows the expected $A^{4/3}$ scaling in the absence of saturation effects. Error bars indicate the $2\sigma$ interval from 25 posterior samples.}
\label{fig:A_dependence}
\end{figure}
Figure~\ref{fig:A_dependence} shows the ratio of the integrated cross section to that for $\gamma+p$ as a function of $A$. 
The results deviate significantly from the $A^{4/3}$ scaling expected without saturation~\cite{Mantysaari:2017slo}, with the suppression increasing monotonically with $A$. 
For the heaviest nuclei, the ratio is reduced to $\sim 0.3$ at $W=31.5$~GeV and to $\sim 0.15$ at $W=813$~GeV. 
The comparison highlights the stronger suppression at higher energies (smaller $x$), leading to about a factor of two difference between the two $W$ values for large $A$, while the two curves approach each other at small $A$.

\section{Conclusion}
We have performed a Bayesian analysis of diffractive $\jpsim$ production in $\gamma+p$ and $\gamma+\mathrm{Pb}$ collisions within the CGC framework. 
While the approach provides a successful description of HERA data, a simultaneous description of proton and nuclear data remains challenging. 
Introducing an overall $K$-factor significantly improves the agreement with experiment by absorbing uncertainties related to the vector meson wave function and higher-order corrections.

Using this framework, we presented predictions for the $A$-dependence of the coherent $\jpsim$ cross section at two photon-nucleus center-of-mass energies. 
The results show a clear departure from the $A^{4/3}$ scaling expected without saturation, with nuclear suppression increasing both with $A$ and with energy. 
For heavy nuclei, the cross section is suppressed by factors of $\sim 0.3$ ($W=31.5$~GeV) and $\sim 0.15$ ($W=813$~GeV) relative to the no-saturation limit.
In the future, a more quantitative assessment of the actual saturation effect will require taking into account the nuclear form factors.
Measuring this $A$ dependence of exclusive vector meson production at the future electron-ion collider (EIC) provides a clean opportunity to investigate the onset of gluon saturation. 
While the energy coverage is modest at the EIC, one can on the other hand explore the dependence of this observable on the photon virtuality $Q^2$.

\section*{Acknowledgements}
H.M. is supported by the Research Council of Finland, the Centre of Excellence in Quark Matter, and projects 338263 and 359902, and under the European Research Council (ERC, grant agreements No. ERC-2023-101123801 GlueSatLight and No. ERC-2018-ADG-835105 YoctoLHC).
This work is supported by the U.S. Department of Energy, Office of Science, Office of Nuclear Physics, under DOE Contract No.~DE-SC0012704 (B.P.S.), DOE Award No. DE-SC0021969 (C.S.) and DE-SC0024232 (C.S. \& H.R.), and within the framework of the Saturated Glue (SURGE) Topical Theory Collaboration (F.S., B.P.S., W.Z.).
H.R. and W.Z. were supported in part by the National Science Foundation (NSF) within the framework of the JETSCAPE collaboration (OAC-2004571).
C.S. acknowledges a DOE Office of Science Early Career Award.
This research was done using resources provided by the Open Science Grid (OSG)~\cite{Pordes:2007zzb,Sfiligoi:2009cct}, which is supported by the NSF awards \#2030508 and \#2323298, and the Wayne State Grid. 
F.S. is supported by the Laboratory Directed Research and Development of Brookhaven National Laboratory and RIKEN-BNL Research Center. 
Part of this work was conducted while F.S. was supported by the Institute for Nuclear Theory of the U.S. DOE under Grant No. DE-FG02-00ER41132. 
The content of this article does not reflect the official opinion of the European Union, and responsibility for the information and views expressed therein lies entirely with the authors.

\bibliography{bib}

\begin{thebibliography}{17}

\bibitem{Gribov:1983ivg}
L.V. Gribov, E.M. Levin, M.G. Ryskin, {Semihard Processes in QCD}, Phys. Rept. \textbf{100}, 1 (1983). \doiwoc{10.1016/0370-1573(83)90022-4}

\bibitem{McLerran:1993ni}
L.D. McLerran, R.~Venugopalan, {Computing quark and gluon distribution functions for very large nuclei}, Phys. Rev. D \textbf{49}, 2233 (1994), \texttt{hep-ph/9309289}. \doiwoc{10.1103/PhysRevD.49.2233}

\bibitem{McLerran:1993ka}
L.D. McLerran, R.~Venugopalan, {Gluon distribution functions for very large nuclei at small transverse momentum}, Phys. Rev. D \textbf{49}, 3352 (1994), \texttt{hep-ph/9311205}. \doiwoc{10.1103/PhysRevD.49.3352}

\bibitem{Ryskin:1992ui}
M.G. Ryskin, {Diffractive $\mathrm{J}/\psi$ electroproduction in LLA QCD}, Z. Phys. C \textbf{57}, 89 (1993). \doiwoc{10.1007/BF01555742}

\bibitem{Mantysaari:2020axf}
H.~M{\"a}ntysaari, {Review of proton and nuclear shape fluctuations at high energy}, Rept. Prog. Phys. \textbf{83}, 082201 (2020), \texttt{2001.10705}. \doiwoc{10.1088/1361-6633/aba347}

\bibitem{Eskola:2022vpi}
K.J. Eskola, C.A. Flett, V.~Guzey, T.~L{\"o}yt{\"a}inen, H.~Paukkunen, {Exclusive J/{\ensuremath{\psi}} photoproduction in ultraperipheral Pb+Pb collisions at the CERN Large Hadron Collider calculated at next-to-leading order perturbative QCD}, Phys. Rev. C \textbf{106}, 035202 (2022), \texttt{2203.11613}. \doiwoc{10.1103/PhysRevC.106.035202}

\bibitem{Lappi:2013am}
T.~Lappi, H.~Mantysaari, {$\mathrm{J}/\psi$ production in ultraperipheral Pb+Pb and $p$+Pb collisions at energies available at the CERN Large Hadron Collider}, Phys. Rev. C \textbf{87}, 032201 (2013), \texttt{1301.4095}. \doiwoc{10.1103/PhysRevC.87.032201}

\bibitem{Mantysaari:2022sux}
H.~M{\"a}ntysaari, F.~Salazar, B.~Schenke, {Nuclear geometry at high energy from exclusive vector meson production}, Phys. Rev. D \textbf{106}, 074019 (2022), \texttt{2207.03712}. \doiwoc{10.1103/PhysRevD.106.074019}

\bibitem{Mantysaari:2016ykx}
H.~M{\"a}ntysaari, B.~Schenke, {Evidence of strong proton shape fluctuations from incoherent diffraction}, Phys. Rev. Lett. \textbf{117}, 052301 (2016), \texttt{1603.04349}. \doiwoc{10.1103/PhysRevLett.117.052301}

\bibitem{Mantysaari:2018zdd}
H.~M{\"a}ntysaari, B.~Schenke, {Confronting impact parameter dependent JIMWLK evolution with HERA data}, Phys. Rev. D \textbf{98}, 034013 (2018), \texttt{1806.06783}. \doiwoc{10.1103/PhysRevD.98.034013}

\bibitem{ALICE:2023jgu}
S.~Acharya et~al. (ALICE), {Energy dependence of coherent photonuclear production of J/{\ensuremath{\psi}} mesons in ultra-peripheral Pb-Pb collisions at $ \sqrt{{\textrm{s}}_{\textrm{NN}}} $ = 5.02 TeV}, JHEP \textbf{10}, 119 (2023), \texttt{2305.19060}. \doiwoc{10.1007/JHEP10(2023)119}

\bibitem{CMS:2023snh}
A.~Tumasyan et~al. (CMS), {Probing Small Bjorken-x Nuclear Gluonic Structure via Coherent J/{\ensuremath{\psi}} Photoproduction in Ultraperipheral Pb-Pb Collisions at sNN=5.02{\,}{\,}TeV}, Phys. Rev. Lett. \textbf{131}, 262301 (2023), \texttt{2303.16984}. \doiwoc{10.1103/PhysRevLett.131.262301}

\bibitem{Mantysaari:2025ltq}
H.~M{\"a}ntysaari, H.~Roch, F.~Salazar, B.~Schenke, C.~Shen, W.~Zhao, {Global Bayesian Analysis of $\mathrm{J}/\psi$ Photoproduction on Proton and Lead Targets} (2025), \texttt{2507.14087}.

\bibitem{Mueller:2001uk}
A.H. Mueller, {A Simple derivation of the JIMWLK equation}, Phys. Lett. B \textbf{523}, 243 (2001), \texttt{hep-ph/0110169}. \doiwoc{10.1016/S0370-2693(01)01343-0}

\bibitem{Mantysaari:2017slo}
H.~M{\"a}ntysaari, R.~Venugopalan, {Systematics of strong nuclear amplification of gluon saturation from exclusive vector meson production in high energy electron{\textendash}nucleus collisions}, Phys. Lett. B \textbf{781}, 664 (2018), \texttt{1712.02508}. \doiwoc{10.1016/j.physletb.2018.04.044}

\bibitem{Pordes:2007zzb}
R.~Pordes et~al., {The Open Science Grid}, J. Phys. Conf. Ser. \textbf{78}, 012057 (2007). \doiwoc{10.1088/1742-6596/78/1/012057}

\bibitem{Sfiligoi:2009cct}
I.~Sfiligoi, D.C. Bradley, B.~Holzman, P.~Mhashilkar, S.~Padhi, F.~Wurthwrin, {The pilot way to Grid resources using glideinWMS}, WRI World Congress \textbf{2}, 428 (2009). \doiwoc{10.1109/CSIE.2009.950}

\end{thebibliography}

\end{document}